\documentclass{article}

\usepackage{titlesec}

\usepackage{PRIMEarxiv}
\usepackage{float}
\usepackage{titlesec}
\usepackage{caption}
\usepackage{PRIMEarxiv}
\usepackage[utf8]{inputenc} 
\usepackage[T1]{fontenc}    
\usepackage{hyperref}       
\usepackage{url}            
\usepackage{booktabs}       
\usepackage{amsfonts}       
\usepackage{nicefrac}       
\usepackage{microtype}      
\usepackage{lipsum}
\usepackage{fancyhdr}       
\usepackage{graphicx}       
\graphicspath{{media/}}
\usepackage{graphicx} 
\usepackage{float} 
\usepackage[
    backend=biber,
    style=numeric,
    sorting=none
  ]{biblatex}

\usepackage{biblatex} 
\usepackage[utf8]{inputenc} 
\usepackage[T1]{fontenc}    
\usepackage{hyperref}       
\usepackage{url}            
\usepackage{booktabs}       
\usepackage{amsfonts}       
\usepackage{nicefrac}       
\usepackage{microtype}      
\usepackage{lipsum}
\usepackage{fancyhdr}       
\usepackage{graphicx}       
\graphicspath{{media/}}     

\title{A Dashboard Approach to Monitoring Mpox-Related Discourse and Misinformation on Social Media

}

\author{
  Linfeng (Leon) Zhao \\
  Computational Media Lab\\
  The University of Texas at Austin \\
  \texttt{leonzhao444@gmail.com} \\
  \And
  Rishul Bhuvanagiri \\
  Computational Media Lab\\
  The University of Texas at Austin \\
  \texttt{rishbhu20@gmail.com} \\  
   \And
  Blake Gonzales \\
  Computational Media Lab\\
  The University of Texas at Austin \\
  \texttt{bng2298@my.utexas.edu} \\  
   \And
  Kellen Sharp \\
  Computational Media Lab\\
  The University of Texas at Austin \\
  \texttt{knsharp@utexas.edu} \\  
    \And
  Dhiraj Murthy \\
  School of Journalism and Media\\
  The University of Texas at Austin \\
  \texttt{dhiraj.murthy@austin.utexas.edu} \\  
}

\begin{document}
\maketitle

\begin{abstract}
Mpox (formerly monkeypox) is a zoonotic disease caused by an orthopoxvirus closely related to variola and remains a significant global public health concern. During outbreaks, social media platforms like X (formerly Twitter) can both inform and misinform the public, complicating efforts to convey accurate health information. To support local response efforts, we developed a researcher-focused dashboard for use by public health stakeholders and the public that enables searching and visualizing mpox-related tweets through an interactive interface. Following the CDC’s designation of mpox as an emerging virus in August 2024, our dashboard recorded a marked increase in tweet volume compared to 2023, illustrating the rapid spread of health discourse across digital platforms. These findings underscore the continued need for real-time social media monitoring tools to support public health communication and track evolving sentiment and misinformation trends at the local level.\end{abstract}


\section{Introduction}
After global concerns first surfaced in summer 2024, the mpox virus has re-emerged as a pressing global health concern. Mpox is a zoonotic orthopoxvirus whose clinical presentation closely resembles smallpox \cite{Intro}. Although initially identified in laboratory monkeys in 1958, the first human case was documented in the Democratic Republic of the Congo in 1970 \cite{who_mpox_}. The virus spreads primarily through direct contact, manifesting initially with flu-like symptoms followed by a blister-like rash, which may appear on the face, hands, genitals, or mouth \cite{cdc_mpox_symptoms}. Since its emergence, mpox has periodically resurfaced, most notably during the 2022 multinational outbreak linked to sub-clade IIb, which resulted in over 100,000 confirmed cases globally \cite{cdc_mpox_symptoms}. In response, the World Health Organization has called for increased international coordination and enhanced disease surveillance systems \cite{who_mpox_}.

However, like the COVID-19 pandemic, managing the biological threat of mpox is only half the battle. Given the public’s reliance on social media to access health information, there are complex health information challenges in a digitally mediated world. Public health systems must therefore also be equipped to detect and respond to infodemics — outbreaks of misinformation, disinformation, information overload, or information voids \cite{chiou}. For example, in the United States, this challenge is amplified by increasing reliance on social media for health information, alongside declining trust in traditional media and government institutions \cite{Chou}. Low health literacy further compounds this vulnerability. As previous studies note, many individuals lack the critical skills needed to evaluate online health content, making them more susceptible to misinformation \cite{paakkari}. Moreover, studies have shown that false information often spreads more quickly and virally than factual content, making it especially difficult to contain in real time \cite{halpern,safarnejad,kolluri2022poxverifiinformationverificationcombat}.

Against this dual crisis of actual viral contagion and the decline of online information integrity, digital dashboards emerge as a promising intervention. These tools not only visualize epidemiological data but also serve as public-facing platforms for combating health misinformation. During the Ebola outbreak, for example, an app integrated with Tableau helped contact tracers in Guinea respond more effectively \cite{Sacks646}. Building on such precedents, recent research has leveraged machine learning and natural language processing to monitor and classify mpox-related discourse on social media. Previous studies demonstrate the potential of AI models to identify credible versus misleading information with over 96 percent accuracy, laying a foundation for more adaptive, data-driven public health communication strategies \cite{kolluri2022poxverifiinformationverificationcombat}.

\section{Research Questions}
In designing our mpox-focused dashboard, we aimed to merge technical precision with cultural awareness. The goal was to create a tool that could support local public health professionals in monitoring online discourse and developing targeted, community-specific communication strategies. The following research questions guided our project:

\begin{enumerate}
    \item[\textbf{RQ1:}] What design principles and technical components make an interactive dashboard a useful decision-support tool for public health agencies?
    \item[\textbf{RQ2:}] How does contemporary social-network culture shape the creation and spread of (mis)information about emerging health threats such as mpox?

\end{enumerate}
\label{sec:headings}

\section{Literature Review}

\subsection{Health Misinformation, Sentiment, and Response Strategies on Social Media}

The 2022 mpox outbreak highlighted the urgency of addressing digital misinformation in real time. Though mpox has a relatively low global case-fatality ratio of 0.17\% and primarily presents as a painful rash \cite{who_mpox_}, public confusion was intensified by the absence of FDA-approved at-home testing in the United States \cite{Perkili}. Computational research methods have become crucial in documenting these information trends online. Previous studies found that sensational and misinformative posts about mpox began to circulate on Twitter well before the WHO declared a Public Health Emergency of International Concern (PHEIC) \cite{Edinger}. In contrast, factual content gained visibility only after this formal recognition.

Subsequent studies reinforce this dynamic. Previous work found that nearly half of 61,000 tweets related to mpox and COVID-19 expressed negative sentiment, reflecting widespread distrust and uncertainty \cite{thakur}. Another study showed that mpox-specific vaccine hesitancy models, fine-tuned with domain-specific training data, outperformed generic RoBERTa models by 8\% in F$_1$ score \cite{perikli}. Similarly, one study found a correlation between spikes in Wikipedia traffic and the creation of new pages, suggesting a feedback loop in public information-seeking behavior, emphasizing the need for real-time monitoring systems that can differentiate reliable guidance from harmful misinformation \cite{ciampaglia}.

\subsection{User Interaction with Misinformation Content}

Platform-level interventions — such as content labeling and removal — offer partial mitigation but often fail to scale, allowing vast amounts of misleading content to go unchecked. To supplement these automated efforts, previous studies advocate for “observational corrections,” in which users directly challenge misinformation \cite{Bode}. While experimental studies show that these peer-led interventions can reduce belief in false claims, maintaining user engagement at scale remains a major obstacle. 

Understanding how users interact with digital content is crucial to designing effective interventions. Previous work categorizes computational defenses into three key strategies: automated fact-checking, source-credibility evaluation, and user-facing interface tools  \cite{karduni}. Vulnerability to misinformation has been found to be rooted in distrust of mainstream media and prolonged exposure to ideological “echo chambers” \cite{Bakshy}. Results from previous work indicate that automated systems alone are insufficient  and that fact-checks must be contextually and culturally meaningful to change behavior \cite{graves}. 

Moreover, user engagement is not always rational or linear. A study found that many users rejected fact-checking widgets, instead relying on personal heuristics and community endorsements to assess credibility \cite{Flintham}. In response, developers have begun designing exploratory tools that accommodate this variability. For example, another study allows users to visualize how specific claims propagate across networks, providing context, related posts, and timelines \cite{Wang}. These tools mark a shift toward consumer empowerment, suggesting a hybrid approach involving computing, psychology, and public education, could enhance the effectiveness of misinformation interventions.

\subsection{Dashboards for Visualizing Health Misinformation}

Dashboards can transform fragmented social media signals into actionable insights. Previous studies developed a self-updating dashboard that analyzed Austrian public sentiment toward COVID-19 policies \cite{Pellert}. Using Twitter data and a national news forum, they mapped keywords to sentiment categories using the LIWC2015 lexicon \cite{Pennebaker} and visualized trends through word clouds and time-series graphs via an R Flexdashboard interface \cite{iannone2023}. Their findings revealed increasing public anxiety over time and demonstrated how accessible visualizations can assist both citizens and policymakers in navigating information overload. 

Other dashboards have fulfilled similar roles during health crises. The Austrian Ministry of Health released a public COVID-19 tracker in March 2020, while the Johns Hopkins Center for Systems Science and Engineering launched its global COVID-19 map the same year \cite{JHU_dashboard}. These platforms provided near-real-time updates, curbing speculation and supporting coordinated responses. Another study developed a dashboard to track COVID-19 clinical trials, helping reduce redundancy and inform research strategy \cite{Thorlund}.

Despite their promise, dashboards face technical constraints. Integrating diverse data sources, maintaining performance, and ensuring user-friendliness remain ongoing challenges. To remain effective, dashboard designs must balance analytical rigor with intuitive visualizations that speak to both experts and the general public. In this context, misinformation-monitoring dashboards represent a critical tool for real-time public health interventions, providing one means to translate data into insights, and insight into action.

\vspace{-0.5em}
\begin{figure}[H]
    \centering
    \includegraphics[width=0.55\textwidth]{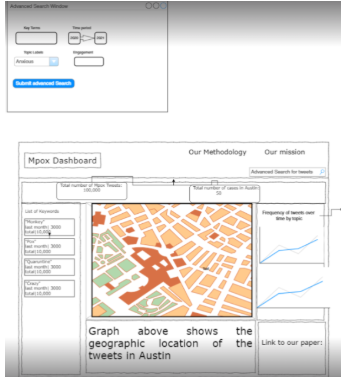} 
    \caption{Wireframe mock-up of the Mpox Dashboard interface.}
\end{figure}
\raggedbottom  

\vspace{-0.5em} 
\section{Methodology}
This project began with the design of the dashboard (see Figure 1) using the diagramming tool Draw.io. This software allowed us to sketch the layout and plan the core features users would interact with. To ensure the dashboard offered a comprehensive view of public discourse, we used three distinct datasets capturing a wide range of mpox-related tweets.

The first dataset that we used was collected by the UT Austin Computational Media Lab (CML) using the Twitter "Spritzer Stream", an API-based random sample of approximately 1\% of global tweets. Because most of the tweets were unrelated to mpox, we developed a Python script to filter for relevant content using keywords such as “mpox” and “Mpox.” Publicly available datasets were largely unusable due to being unhydrated. To address this, we obtained permission to use a hydrated, pre-filtered “Monkeypox dataset” from May 2022 \cite{Nia_dataset}. 

As the Twitter API is mostly limited to paid access, we used alternative tools to collect recent tweets. We used Zeeschuimer \cite{zeeschuimer2023}, a research-oriented social media data collection tool, to collect 3,000 tweets from 2024 related to mpox. These were integrated with the CML and “Monkeypox dataset” \cite{Nia_dataset} to together form a more comprehensive dataset for analysis.

After collecting the datasets, we ran a Python-based filtering pipeline in Visual Studio Code to extract and consolidate mpox-related content. The script used the pandas library to process CSV files and isolate relevant tweets, with the os library handling file management tasks. This process yielded a clean dataset focused solely on mpox discourse, which was then visualized through the dashboard. After finalizing our dataset, we developed the dashboard using Streamlit, an open-source web application framework. We chose Streamlit for its flexibility and efficiency, as it enabled us to present data in multiple formats such as searchable tables, interactive graphs, and real-time visualizations. These features made the final product highly accessible and user-friendly for public health researchers.

\section{Results}
\subsection{Dashboard Design}

Our dashboard is designed to organize mpox-related tweets and visualize them in ways most useful to public health professionals and researchers (see Figure 2). While other mpox-related tools like  the Poxverifi tool \cite{kolluri2022poxverifiinformationverificationcombat} are intended for general public use, our platform is tailored to support organizations like local public health organizations by helping them understand mpox discourse and develop targeted health literacy campaigns to counter misinformation. The design draws inspiration from the Johns Hopkins COVID-19 dashboard \cite{JHU_dashboard}, which successfully visualized real-time data on COVID-19 cases, deaths, and locations.

In alignment with most local public health organizations' goals — e.g., prevention, community self-sufficiency, and cross-sector collaboration —the dashboard provides searchable tables and visualizations showing tweet locations and keyword trends over time. These tools help identify emerging concerns and guide strategic communication efforts.

A key feature is the advanced search function, which allows users to input up to three keywords and filter by engagement metrics such as like count, reply count, and retweet count. This enables more focused analysis and supports efforts to track the spread of misinformation. By surfacing engagement patterns, the dashboard also offers insights into content-consumer dynamics, highlighting which narratives gain traction and why.

\begin{figure}[H]
    \centering
    \includegraphics[width=0.7\textwidth]{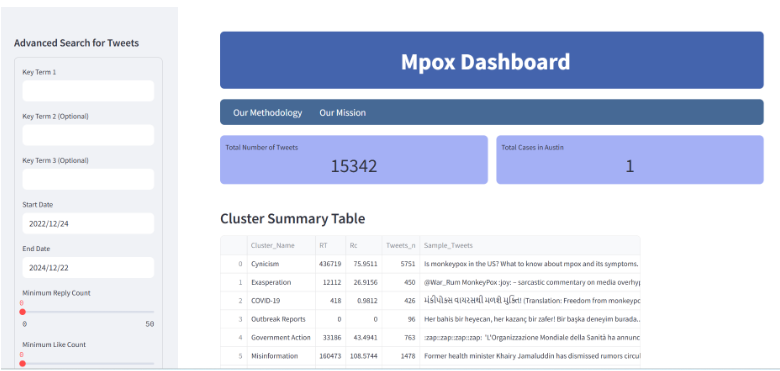} 
    \caption{Front page of the mpox dashboard.}
    
\end{figure}

\subsection{Sentiment and Topic Clustering Over Time}

To enable comparisons over time, we applied an algorithm to label tweets according to their primary topics and sentiment, then visualized the results using a color-coded cluster graph (see Figure 3). The resulting scatterplot illustrates the relative frequency of key thematic clusters such as cynicism, COVID-19 comparisons, government action, and misinformation over time. Each point represents the proportion of tweets in a specific category on a given day.
The graph shows that cynicism became the dominant sentiment in mpox-related discussions, particularly in the more recent data. This trend may reflect the virus’s increased spread and growing media coverage. Cynicism, defined by distrust in public health institutions and traditional news sources, can make the public more susceptible to unofficial or misleading health information.
To support targeted analysis, we also developed an advanced search feature that allows researchers to filter tweets by specific keywords and time frames. This addition makes the user interface more flexible and better suited to the diverse needs of public health researchers.

\begin{figure}[H]
    \centering
    \includegraphics[width=0.8\textwidth]{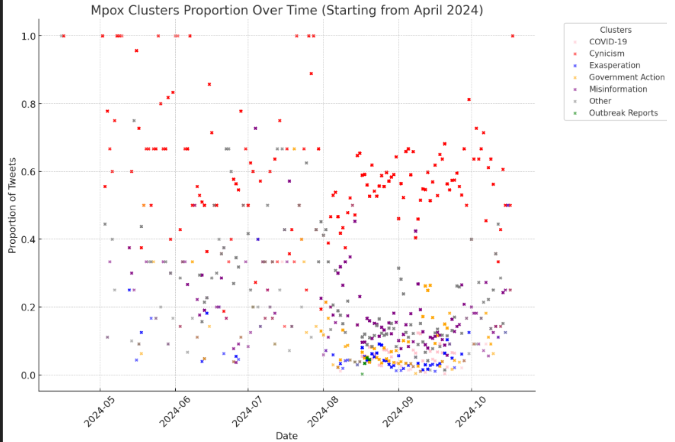} 
    \caption{Mpox tweet cluster proportions over time (starting April 2024).}
\end{figure}

\vspace{-1em}

\section{Future Work and Limitations}

While this study provides valuable insights into mpox-related discourse on Twitter, it has several limitations that point to opportunities for future research. A primary limitation is the dataset’s timeframe, which spans only 2023 to 2024. This narrow window may not fully reflect the evolution of public sentiment, awareness, or misinformation over time. To build on these findings, future studies should consider collecting data over a longer period to better capture shifting narratives and reactions to new developments. Expanding beyond X (formerly Twitter) to include Facebook, Instagram, TikTok, and Reddit would also enhance the analysis. These platforms serve different user bases and support varied forms of communication, offering richer insights into how diverse communities engage with public health information. By broadening both the temporal and platform scope of analysis, future research can develop a more comprehensive understanding of how social media ecosystems shape the creation, spread, and reception of (mis)information during public health crises.

\section{Conclusion}

In an era where public health crises are compounded by waves of digital misinformation, our mpox-focused dashboard offers a researcher-centered tool for translating social media discourse into actionable insight. Designed with local public health agencies in mind, the dashboard moves beyond reactive fact-checking by foregrounding visibility into real-time sentiment, engagement, and thematic trends. Our analysis revealed the growing prominence of cynicism and misinformation in mpox-related conversations, underscoring the role of platform culture in shaping public understanding. By integrating advanced search, keyword filtering, and engagement metrics, the dashboard enables researchers to identify high-impact narratives and track evolving discourse. In doing so, it bridges the gap between public health data and the digital environments in which that data circulates, supporting more targeted, culturally informed communication strategies.

\printbibliography

@article{Intro,
title = {Unveiling the Mpox menace: exploring the intricacies of a zoonotic virus and clinical implications},
journal = {Diagnostic Microbiology and Infectious Disease},
volume = {107},
number = {2},
pages = {116024},
year = {2023},
issn = {0732-8893},
doi = {https://doi.org/10.1016/j.diagmicrobio.2023.116024},
url = {https://www.sciencedirect.com/science/article/pii/S0732889323001347},
author = {Eshita Sharma and Sakshi Malhotra and Shreya Kaul and Neha Jain and Upendra Nagaich}
}

@article{chiou,
  title={The future of infodemic surveillance as public health surveillance},
  author={Chiou, Howard and Voegeli, Christopher and Wilhelm, Elisabeth and Kolis, Jessica and Brookmeyer, Kathryn and Prybylski, Dimitri},
  journal={Emerging Infectious Diseases},
  volume={28},
  number={Suppl 1},
  pages={S121},
  year={2022}
}

@article {Sacks646,
	author = {Sacks, Jilian A and Zehe, Elizabeth and Redick, Cindil and Bah, Alhoussaine and Cowger, Kai and Camara, Mamady and Diallo, Aboubacar and Gigo, Abdel Nasser Iro and Dhillon, Ranu S and Liu, Anne},
	title = {Introduction of Mobile Health Tools to Support Ebola Surveillance and Contact Tracing in Guinea},
	volume = {3},
	number = {4},
	pages = {646--659},
	year = {2015},
	doi = {10.9745/GHSP-D-15-00207},
	publisher = {Global Health: Science and Practice},
	URL = {https://www.ghspjournal.org/content/3/4/646},
	eprint = {https://www.ghspjournal.org/content/3/4/646.full.pdf},
	journal = {Global Health: Science and Practice}
}

@misc{kolluri2022poxverifiinformationverificationcombat,
      title={PoxVerifi: An Information Verification System to Combat Monkeypox Misinformation}, 
      author={Akaash Kolluri and Kami Vinton and Dhiraj Murthy},
      year={2022},
      eprint={2209.09300},
      archivePrefix={arXiv},
      primaryClass={cs.CL},
      url={https://arxiv.org/abs/2209.09300}, 
}

@article{Chou,
author = {Sylvia Chou, Wen-Ying and Gaysynsky, Anna},
title = {A Prologue to the Special Issue: Health Misinformation on Social Media},
journal = {American Journal of Public Health},
volume = {110},
number = {S3},
pages = {S270-S272},
year = {2020},
doi = {10.2105/AJPH.2020.305943},
    note ={PMID: 33001727},

URL = { 
    
        https://doi.org/10.2105/AJPH.2020.305943
},
eprint = {    
        https://doi.org/10.2105/AJPH.2020.305943
}
}

@inproceedings{halpern,
  title={From belief in conspiracy theories to trust in others: Which factors influence exposure, believing and sharing fake news},
  author={Halpern, Daniel and Valenzuela, Sebasti{\'a}n and Katz, James and Miranda, Juan Pablo},
  booktitle={Social Computing and Social Media. Design, Human Behavior and Analytics: 11th International Conference, SCSM 2019, Held as Part of the 21st HCI International Conference, HCII 2019, Orlando, FL, USA, July 26-31, 2019, Proceedings, Part I 21},
  pages={217--232},
  year={2019},
  organization={Springer}
}

@article{safarnejad,
  title={Contrasting misinformation and real-information dissemination network structures on social media during a health emergency},
  author={Safarnejad, Lida and Xu, Qian and Ge, Yaorong and Krishnan, Siddharth and Bagarvathi, Arunkumar and Chen, Shi},
  journal={American journal of public health},
  volume={110},
  number={S3},
  pages={S340--S347},
  year={2020},
  publisher={American Public Health Association}
}

@article{perikli,
  title={Detecting the presence of COVID-19 vaccination hesitancy from South African twitter data using machine learning},
  author={Perikli, Nicholas and Bhattacharya, Srimoy and Ogbuokiri, Blessing and Nia, Zahra Movahedi and Lieberman, Benjamin and Tripathi, Nidhi and Dahbi, Salah-Eddine and Stevenson, Finn and Bragazzi, Nicola and Kong, Jude and others},
  journal={arXiv preprint arXiv:2307.15072},
  year={2023}
}

@article{edinger,
  title={Misinformation and public health messaging in the early stages of the mpox outbreak: mapping the Twitter narrative with deep learning},
  author={Edinger, Andy and Valdez, Danny and Walsh-Buhi, Eric and Trueblood, Jennifer S and Lorenzo-Luaces, Lorenzo and Rutter, Lauren A and Bollen, Johan},
  journal={Journal of Medical Internet Research},
  volume={25},
  pages={e43841},
  year={2023},
  publisher={JMIR Publications Toronto, Canada}
}

@article{thakur,
  title={Sentiment analysis and text analysis of the public discourse on Twitter about COVID-19 and MPox},
  author={Thakur, Nirmalya},
  journal={Big Data and Cognitive Computing},
  volume={7},
  number={2},
  pages={116},
  year={2023},
  publisher={MDPI}
}

@article{ciampaglia,
  title={The production of information in the attention economy},
  author={Ciampaglia, Giovanni Luca and Flammini, Alessandro and Menczer, Filippo},
  journal={Scientific reports},
  volume={5},
  number={1},
  pages={9452},
  year={2015},
  publisher={Nature Publishing Group UK London}
}

@article{bode,
  title={User correction as a tool in the battle against social media misinformation},
  author={Bode, Leticia},
  journal={Geo. L. Tech. Rev.},
  volume={4},
  pages={367},
  year={2019},
  publisher={HeinOnline}
}

@article{paakkari,
  title={COVID-19: health literacy is an underestimated problem},
  author={Paakkari, Leena and Okan, Orkan},
  journal={The lancet public health},
  volume={5},
  number={5},
  pages={e249--e250},
  year={2020},
  publisher={Elsevier}
}

@article{graves,
  title={Understanding the promise and limits of automated fact-checking},
  author={Graves, D},
  journal={Reuters Institute for the Study of Journalism},
  year={2018},
  publisher={Reuters Institute for the Study of Journalism}
}

@article {Nia_dataset,
	Title = {A Twitter dataset for Monkeypox, May 2022},
	Author = {Nia, Zahra M and Bragazzi, Nicola L and Wu, Jianhong and Kong, Jude D},
	DOI = {10.1016/j.dib.2023.109118},
	Volume = {48},
	Month = {June},
	Year = {2023},
	Journal = {Data in brief},
	ISSN = {2352-3409},
	Pages = {109118},
	URL = {https://europepmc.org/articles/PMC10102531},
}

@misc{zeeschuimer2023,
  author       = {{Digital Methods Initiative}},
  title        = {Zeeschuimer},
  year         = {2023},
  howpublished = {\url{https://github.com/orgs/digitalmethodsinitiative/repositories}},
  note         = {GitHub repository},
}

@misc{iannone2023,
  author       = {Iannone, J. and Allaire, J. J. and Borges, B. and Sievert, C.},
  title        = {flexdashboard: R Markdown format for flexible dashboards},
  year         = {2023},
  howpublished = {\url{https://pkgs.rstudio.com/flexdashboard/}},
  note         = {R package version 0.6.1, RStudio},
}

@article{JHU_dashboard,
author = {Dong, Ensheng and Ratcliff, Jeremy and Goyea, Tamara and Katz, Aaron and Lau, Ryan and Ng, Timothy and Garcia, Beatrice and Bolt, Evan and Prata, Sarah and Zhang, David and Murray, Reina and Blake, Mara and Du, Hongru and Ganjkhanloo, Fardin and Ahmadi, Farzin and Williams, Jason and Choudhury, Sayeed and Gardner, Lauren},
year = {2022},
month = {08},
pages = {},
title = {The Johns Hopkins University Center for Systems Science and Engineering COVID-19 Dashboard: data collection process, challenges faced, and lessons learned},
volume = {22},
journal = {The Lancet Infectious Diseases},
doi = {10.1016/S1473-3099(22)00434-0}
}

@article{Thorlund,
author = {Thorlund, Kristian and Dron, Louis and Park, Jay and Hsu, Grace and Forrest, Jamie and Mills, Edward},
year = {2020},
month = {04},
pages = {},
title = {A real-time dashboard of clinical trials for COVID-19},
volume = {2},
journal = {The Lancet Digital Health},
doi = {10.1016/S2589-7500(20)30086-8}
}

@inproceedings{Wang,
   title={RumorLens: Interactive Analysis and Validation of Suspected Rumors on Social Media},
   url={http://dx.doi.org/10.1145/3491101.3519712},
   DOI={10.1145/3491101.3519712},
   booktitle={CHI Conference on Human Factors in Computing Systems Extended Abstracts},
   publisher={ACM},
   author={Wang, Ran and Du, Kehan and Chen, Qianhe and Zhao, Yifei and Tang, Mojie and Tao, Hongxi and Wang, Shipan and Li, Yiyao and Wang, Yong},
   year={2022},
   month=apr, pages={1–7},
   collection={CHI ’22} }

@inproceedings{Flintham,
author = {Flintham, Martin and Karner, Christian and Bachour, Khaled and Creswick, Helen and Gupta, Neha and Moran, Stuart},
title = {Falling for Fake News: Investigating the Consumption of News via Social Media},
year = {2018},
isbn = {9781450356206},
publisher = {Association for Computing Machinery},
address = {New York, NY, USA},
url = {https://doi.org/10.1145/3173574.3173950},
doi = {10.1145/3173574.3173950},
booktitle = {Proceedings of the 2018 CHI Conference on Human Factors in Computing Systems},
pages = {1–10},
numpages = {10},
location = {Montreal QC, Canada},
series = {CHI '18}
}

@article{Bakshy,
author = {Bakshy, Eytan and Messing, Solomon and Adamic, Lada},
year = {2015},
month = {05},
pages = {},
title = {Political science. Exposure to ideologically diverse news and opinion on Facebook},
volume = {348},
journal = {Science (New York, N.Y.)},
doi = {10.1126/science.aaa1160}
}

@misc{karduni,
      title={Human-Misinformation interaction: Understanding the interdisciplinary approach needed to computationally combat false information}, 
      author={Alireza Karduni},
      year={2019},
      eprint={1903.07136},
      archivePrefix={arXiv},
      primaryClass={cs.HC},
      url={https://arxiv.org/abs/1903.07136}, 
}

@misc{Perkili,
      title={Detecting the Presence of COVID-19 Vaccination Hesitancy from South African Twitter Data Using Machine Learning}, 
      author={Nicholas Perikli and Srimoy Bhattacharya and Blessing Ogbuokiri and Zahra Movahedi Nia and Benjamin Lieberman and Nidhi Tripathi and Salah-Eddine Dahbi and Finn Stevenson and Nicola Bragazzi and Jude Kong and Bruce Mellado},
      year={2023},
      eprint={2307.15072},
      archivePrefix={arXiv},
      primaryClass={cs.CY},
      url={https://arxiv.org/abs/2307.15072}, 
}

@online{who_mpox_,
  author       = {{World Health Organization}},
  title        = {Mpox},
  year         = {2025},
  url          = {https://www.who.int/news-room/fact-sheets/detail/mpox},
  note         = {Accessed: 2025-05-06},
  organization = {World Health Organization}
}

@online{cdc_mpox_symptoms,
  author       = {{Centers for Disease Control and Prevention}},
  title        = {Signs and Symptoms of Mpox},
  year         = {2024},
  url          = {https://www.cdc.gov/mpox/signs-symptoms/index.html},
  note         = {Accessed: 2025-05-06},
  organization = {Centers for Disease Control and Prevention}
}

@unknown{Pellert,
author = {Pellert, Max and Lasser, Jana and Metzler, Hannah and Garcia, David},
year = {2020},
month = {06},
pages = {},
title = {Dashboard of sentiment in Austrian social media during COVID-19},
doi = {10.48550/arXiv.2006.11158}
}

@unknown{Pennebaker,
author = {Boyd, Ryan and Ashokkumar, Ashwini and Seraj, Sarah and Pennebaker, James},
year = {2022},
month = {02},
pages = {},
title = {The Development and Psychometric Properties of LIWC-22},
doi = {10.13140/RG.2.2.23890.43205}
}

\end{document}